\documentclass[journal=jpccck,manuscript=article]{achemso}
\usepackage{epsfig}
\usepackage{graphicx,epstopdf}
\usepackage{psfrag}
\usepackage{ae}
\usepackage{amsmath,amssymb}
\usepackage[usenames]{color}
\usepackage{float}

\title{Potassium bromide, KBr/$\varepsilon$: New Force Field}

\author{ Ra\'ul Fuentes-Azcatl} 
\email{razcatl@hotmail.com}
\affiliation{Instituto 
de F\'{\i}sica, Universidade Federal
do Rio Grande do Sul, Caixa Postal 15051, CEP 91501-970, 
Porto Alegre, RS, Brazil}
\author{Marcia C. Barbosa} 
\email{marcia.barbosa@ufrgs.br}
\affiliation{Instituto 
de F\'{\i}sica, Universidade Federal
do Rio Grande do Sul, Caixa Postal 15051, CEP 91501-970, 
Porto Alegre, RS, Brazil}

\begin{document}

\date{}

\begin{abstract}
A force field needs to
 reproduce coincident many properties of ions, like their structure, solvation, and moreover
both the interactions of these ions with each other in the crystal and in solution and the interactions of ions
 with other molecules. Using a similar strategy employed in the parameterization of the NaCl/$\varepsilon$~\cite{NaCle}, in this paper, we first propose a
 force field for the Potassium Bromide, the KBr/$\varepsilon$. This
new model is compared with the experimental
values of cristal density and structure for the salt  and the density, the viscosity, the dielectric constant 
and the solubility in the water solution for a range of 
concentrations. Next, the  transferability, of this new model KBr/$\varepsilon$ and the NaCl/$\varepsilon$ , is verified by
creating the KCl/$\varepsilon$ and the NaBr/$\varepsilon$ models.
The strategy is to employ
the same parameters obtained for the NaCl/$\varepsilon$ and for
the KBr/$\varepsilon$ force fields. The two new models
derived are also compared with the experimental
values for the density, the viscosity, the dielectric constant 
and the solubility in the water solution for a range of 
concentrations.
 
\end{abstract}

\maketitle

\section{Introduction}

The potassium Bromide salt shows a number of 
applications in medicine particularly in the metabolic
regulation~\cite{ID}. The potassium levels influence
 multiple physiological processes, including~\cite{ID,Malnic,Mount}
the  cellular-membrane potential, the propagation of action potentials 
in neuronal, in the muscular,
 and in the cardiac tissue. 
Recent studies suggest that bromine is necessary for 
tissue development~\cite{McCall, Mayeno,Patricelli} and is relevant
in the antiparasitic enzyme in the human immune system. 

The salt interaction with the biological 
system is quite complex and experiments even though very
important are unable to isolate the properties
of the individual molecule-molecule interaction. Therefore, one strategy 
to understand the interaction of the  KBr with
other molecules is the use of simulations.
The crucial step in the simulations
is to construct
an appropriated force field for the interaction
potential between the ions. The usual method is to
fit the parameters
of the  model  with the experimental 
results for  the density
and for the structure for the
real system  at one
 determined  pressure
and temperature. Then, the results
obtained for thermodynamic and dynamic properties with the
 model are compared with experiments. Following this procedure, 
atomistic models for KBr have been proposed~\cite{JC08}. Unfortunately,
even though capable of reproducing the density of the system
at 298K and 1bar the current models
fail in reproducing other properties
such as the  the dielectric constant of the 
solution in water, the viscosity and the solubility.

Recently we proposed a new model for Sodium Chloride~\cite{NaCle},
the NaCl/$\varepsilon$,  which
is able to reproduce not only the density, but the dielectric
constant, the viscosity and the solubility of this
salt in aqueous solution, as well as,
 the density of the pure system at different temperatures. 
The idea of adjusting the model to give the experimental
dielectric constant of the pure system and of the water 
solution at 1bar and 298K is inspired by the
recent need to understand the behavior of salt in
surfaces and confined geometries and
in the solubility where the 
dielectric discontinuity plays quite a relevant role~\cite{KF-vega,Corradini,Liu}.

The remaining of the 
paper goes as follows. In the section 2 
the new model, the KBr/$\varepsilon$, is introduced 
 and the TIP4P/$\varepsilon$ water model
was  reviewed. Section 3 summarizes
the simulation details
and  the results are analysed in Section 4. Conclusions
are presented in the section 5.

\section{The Models}

\subsection{The KBr/$\varepsilon$ Model}

The ions of the salt are modeled as spherical particles  
interacting through the potential
\begin{equation}
\label{ff}
u(r_{ij}) = 4\epsilon_{ij} 
\left[\left(\frac {\sigma_{ij}}{r_{ij}}\right)^{12}
-\left (\frac{\sigma_{ij}}{r_{ij}}\right)^6\right] 
+ \lambda_i\lambda_j\frac{q_iq_j}{4\pi\epsilon_0r_{ij}}
\end{equation}
\noindent where $r_{ij}$ is the distance between ions $i$ and $j$, $q_i$ is 
the electric charge of ion $i$, $\epsilon_0$ is the permitivity of 
vacuum,  $\epsilon_{ij}$ is the Lennard-Jones 
energy that is used as energy scale and  $\sigma_{ij}$ is the 
distance between the ions, used as length scale.

For the interaction between the 
ions and the water molecules , the Lorentz-Berteloth 
rule is employed~\cite{Hansen}, namely
\begin{equation}
\label{ff}
\sigma_{\alpha\beta}=\bigg(\frac{\sigma_{\alpha\alpha}
+\sigma_{\beta\beta}}{2}\bigg)\ \ ;\ \ \epsilon_{\alpha\beta}
=\sqrt {\epsilon_{\alpha\alpha}\epsilon_{\beta\beta}}\; .
\end{equation}

For the KBr/$\varepsilon$ model  the Lennard-Jones (LJ) 
energy, $\epsilon_{ij}=\epsilon_{LJ}$,and the distance
scale, $\sigma_{ij}=\sigma_{LJ}$, are the same 
for any $i$ and $j$ namely  K-K, K-Br or Br-Br.
The ion charges are $q_{i}=\pm 1\;e$
where $e$ is the charge of an electron.

Recently a new model for the 
NaCl, the NaCl/$\varepsilon$, was proposed.
In this system, a screening for the Coulombic term 
 was introduced~\cite{St14}. 
The original assumption was to 
include  effects due to water  polarization by adding
to a rigid model a screening term~\cite{St14,NaCle}. Here we addapt
this proposition for the  KBr salt model.
Then 
the screening parameter becomes  $\lambda_i=\lambda_C$.

Therefore, our interaction potential of the ions has
 three parameters, namely $\lambda_{C}$, $\sigma_{LJ}$ and  $\epsilon_{LJ}$
to be adjusted with experimental data for each ion.
These parameters are selected so the  KBr/$\varepsilon$ 
force field  reproduces  the 
 experimental value for the  density 
of the crystal in the face centred cubic phase
at the  $298\; K$ of temperature~\cite{CRC,JC08} and $1\;bar$.
These procedure allows for a number of possible
parameters values. This degeneracy is broken 
by selecting the subset that also gives the
  radial distribution function, g(r), which
gives the appropriate behavior of 
the salt crystal at $298\; K$  of temperature
and $1\;bar$ of pressure.

The next step is this reduction in the
parameter space is to select the set of values that also 
gives the 
proper density and the dielectric constant in the mixture of
the salt with water~\cite{CRC} at $298\; K$  of temperature
and $1\;bar$ of pressure. 
 These last step   was done using a solution
with  $4$ molal of salt concentration and the
TIP4P/$\varepsilon$ water model . The final
result  for 
the force field for the KBr/$\varepsilon$ model 
is shown in the Table~\ref{tab:KBrParam}.

 \begin{table}[h]
\caption{Force field parameter of KBr/$\varepsilon$. }
\label{tab:KBrParam}
\begin{center}
\begin{tabular}{|ccccc|}
\hline\hline
Model  & q/e&$\lambda_{C}$& 
$\sigma$/\AA & $(\epsilon/k_B)$/K\\
\hline
K&+1&0.885&2.86&115.83\\
Br&-1&0.885&4.057&287.47\\
\hline
\end{tabular}
\end{center}
\end{table}
 
\subsection{TIP4P/$\varepsilon$ Water Model}

\begin{figure}
\centerline{\psfig{figure=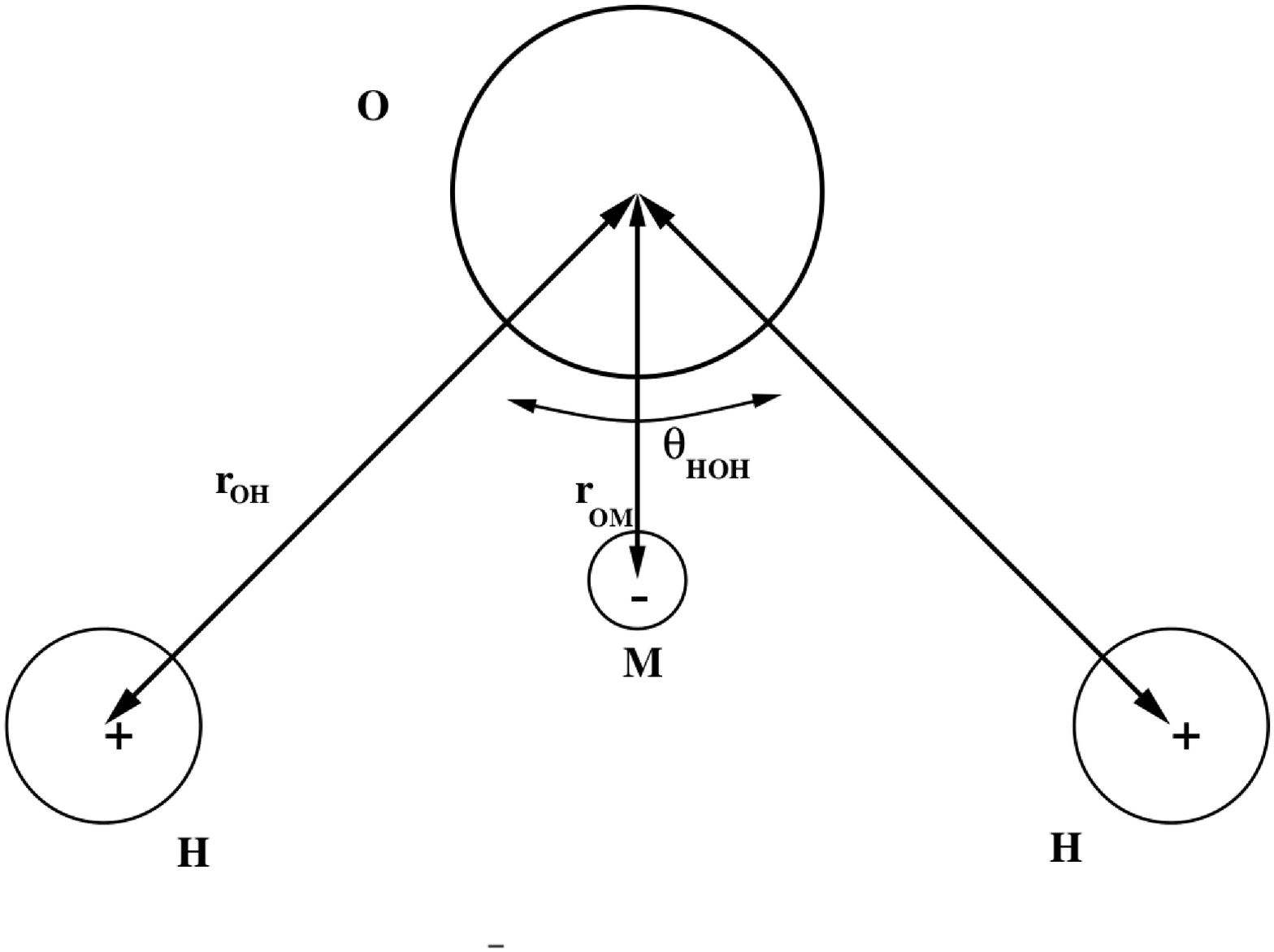,width=10.0cm,angle=0}}
\caption{Geometry forcefield of TIP4P/$\varepsilon$ water, based on the 
geometry of the force field TIP4P\cite{tip4p}. The geometry have a positive 
charges on every nucleus of H and a negative charge at a distance r$_{OM}$ along
 the bisector of the bending angle.}
\label{tip4pe}
\end{figure}

The TIP4P/$\varepsilon$~\cite{tip4pe} water is
illustrated 
in the figure~\ref{tip4pe}. The intermolecular force 
is given by the Lennard Jones  and the Coulomb 
interactions as given
by the Eq.~\ref{ff}. 
The positive charges are located 
at  each hydrogen and 
 the negative charge which neutralizes
the molecule is placed  along the bisector of the HOH 
angle located at distance $l_{OM}$ of the 
oxygen as shown in the figure~\ref{tip4pe}. The parameters 
of the Force Fields for the TIP4P/$\varepsilon$ are  given in the 
Table~\ref{tab:tip4pe}
with  $\lambda_O=\lambda_H=1$ in the Eq.~\ref{ff}.



\begin{table}[h]
\caption{Force field parameters of TIP4P/$\varepsilon$ water model. The 
charge in site $M$ is $q_M=-(2 q_H)$.  }
\label{tab:tip4pe}
\begin{center}
\begin{tabular}[h]{|cccccccc|}
\hline
\hline Model & $r_{OH}$/\AA  & $\Theta$/ $^0$  & $q_H/e$  & $q_M/e$& $r_{OM}$ /\AA  & 
$\sigma$/\AA & $(\epsilon/k_B)$/K\\
\hline

TIP4P/$\varepsilon$ & 0.9572 & 104.52 &0.527& 1.054& 0.105 &  3.165 
& 93\\

\hline
\end{tabular}
\end {center}
\end{table}

\newpage

\section{The Simulation Details}

Molecular dynamic (MD) simulations were performed using 
GROMACS~\cite{gromacs}. The equations of motion were
solved using the leap-frog algorithm~\cite{allen,gromacs} with
$2\; fs$ time steps. The total time for the simulation for
different molalities 
 is $30$  ns, 
keeping the positions and velocities for every 500 steps. 

For the  shear viscosity shorter times steps
and longer simulations were employed,  $1\;fs$ and 
$40$ ns  respectively.
The Coulombic forces were treated via Ewald 
summations with the real part of the Coulombic
 potential truncated at $10$\AA. The Fourier component of the Ewald 
sums was evaluated by using the smooth particle mesh Ewald (SPME) 
method~\cite{SPME} with a grid spacing of $1.2$\AA and a fourth degree
 polynomial for the interpolation. 
The simulation box is cubic 
throughout the whole simulation and the geometry of the water molecules
 kept constant using the LINCS procedure~\cite{LINCS}. The NpT
ensembled was employed with the Nos\'e Hoover thermostat~\cite{nhc}
and the Parinell-Rahman barostat with 
a $\tau_P$ parameter of 1.0 ps~\cite{gromacs}.

The MD simulations for the pure KBr were carried out 
under $1\;bar$ pressure condition, on a system of $1024$ KBr pairs, with a 
time step $\bigtriangleup t = 2\;fs$, the time of simulations is 10 ns 
and storing the positions and velocities every 1000 simulation step. 
For bromide potacium in water, the simulations 
have been done using $864$ molecules  in the
liquid phase at different molalities and at the temperature of $298\;K$ 
and $1\;bar$ of 
pressure. The molality concentration is obtained 
from the total number of ions in solution $N_{ions}$ , the number of 
water molecules $N_{H_{2}O}$ and the molar mass of water $M_{H_{2}O}$ as:\\
 \begin{equation}
\label{mol}
\left[KBr \right] =\frac{N_{ions}\times 10^{3}}{2N_{H_{2}O}M_{H_{2}O}}\; .
\end{equation} \\
The  division by 2 in this equation accounts for a pair of ions 
and $M_{H_{2}O}$ =18 g mol$^{-1}$. The figure~\ref{tab:molal}
 gives the value of the molality for each point of calculus
 \begin{table}[h]
\caption{Composition of KBr solutions used in the simulations  
at $298.15\; K$ and $1\;bar$. }
\label{tab:molal}
\begin{center}
\begin{tabular}{|ccc|}
\hline\hline
Molality (m)  & $N_{H_{2}O}$ & $N_{ions}$ \\
\hline
0.99&832&32\\
1.99&806&58\\
3.07&778&86\\
4.05&754&110\\
5.0&732&132\\

	\hline
\end{tabular}
\end{center}
\end{table}

The static dielectric constant is computed from the 
fluctuations~\cite{neumann} of the total dipole moment {\bf M},
\begin{equation}
\epsilon=1+\frac{4\pi}{3k_BTV} (<{\bf M}^2>-<{\bf M}>^2)
\end{equation}
\noindent where  $k_B$ is the Boltzmann constant and $T$ the absolute 
temperature. The dielectric constant is obtained for long simulations at 
constant density and temperature or at constant temperature and pressure.
The shear viscosity is obtained using the autocorrelation function of the 
off-diagonal components of the pressure tensor $P_{\alpha\beta}$ according 
to the Green-Kubo formulation,
\begin{equation}
\eta= \frac{V}{k_BT} \int_0^\infty <P_{\alpha\beta}(t_0)P_{\alpha\beta}
(t_0+t)>_{t_0}dt,
\end{equation}\\

\section{Results}

\subsection{ The KBr/$\varepsilon$ Model}

\begin{figure}
\centerline{\psfig{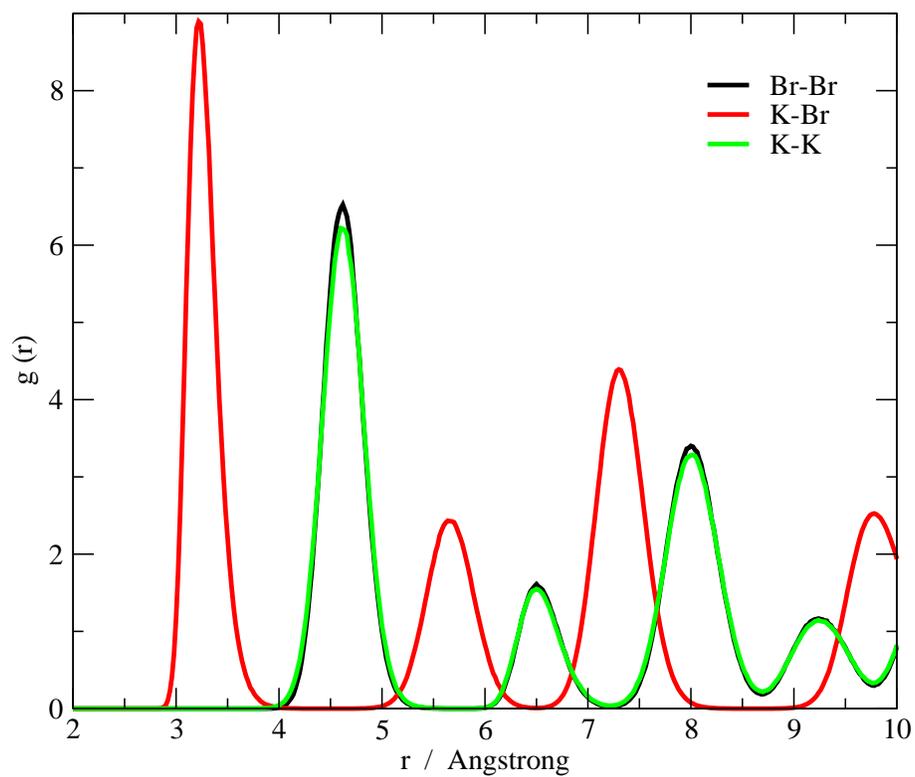}}
\caption{Radial distribution function g(r) versus the distance r at 1
bar and 298 K for: K-Br (red line), K-K (green line), and Br-Br
(black line).
}
\label{Fig-gr-KBr}
\end{figure}

The parameters for KBr/$\varepsilon$ model were
selected to reproduce the   density of the crystal 
at the $298\;K$ and $1 bar$, namely $2.74 \;g \;cm^{-3}$~\cite{CRC}
and the to show the peak in the  radial distribution  at 
$3.29\AA$
given  in the figure~\ref{Fig-gr-KBr}  in agreement with 
the experiments~\cite{CRC}. The final step of the 
model is obtained by adjusting the 
parameters to give the correct dielectric
constant of the  4 molal solution of the salt in
water.
 \\
 \\

The Lattice Energy (LE) of the  KBr/$\varepsilon$ model is
 $582.9\;kJ/mol$ while  the  Lattice
Constant (LC) is $6.58\;\AA$ what is 
comparable with 
the experimental data for these
two quantities  that are given 
by  $671.11\;kJ/mol$~\cite{CRC} and 
by $6.6\;\AA$~\cite{CRC} respectively.

The Table~\ref{tab:FFcompKBr} shows a comparison of the values obtained 
for the
the density, the Lattice Energy and the Lattice  Crystal 
for the  KBr/$\varepsilon$ model, for the experiments~\cite{CRC},  for  the  
Joung-Cheatham~\cite{JC08}, the \textbf{JC} force field, and for the force 
field parametrized
with SPC/E water, the JC$_{S3}$ model~\cite{SD94}.
 Our model gives
good agreement with the experiments~\cite{CRC} in the density of the 
crystal and the 
Lattice Constant, but
is 13$\%$ far from the reproduction of the Lattice Energy.

\begin{table}[h]
\caption{Density of KBr at 1 bar of pressure and 298 K of temperature,
Lattice Energy, Lattice  Crystal 
of various Force Fields and for experiments~\cite{CRC}.}
\label{tab:FFcompKBr}
\begin{center}
\begin{tabular}{|cccc|}
\hline
 Model Ions& $\rho/(g/cm^3)$ & LC/\AA&LE/(kJ/mol) \\
\hline
JC$_{S3}$~\cite{SD94}& 2.61 &6.66&695.38\\
JC$_{T4}$~\cite{JC08} &2.67&6.62 &698.72\\

 KBr/$\varepsilon$ & 2.76& 6.58&582.9\\
experimental~\cite{CRC}  &2.74&6.6&671.11\\
\hline
\end{tabular}
\end{center}
\end{table}

\begin{figure}
\centerline{\psfig{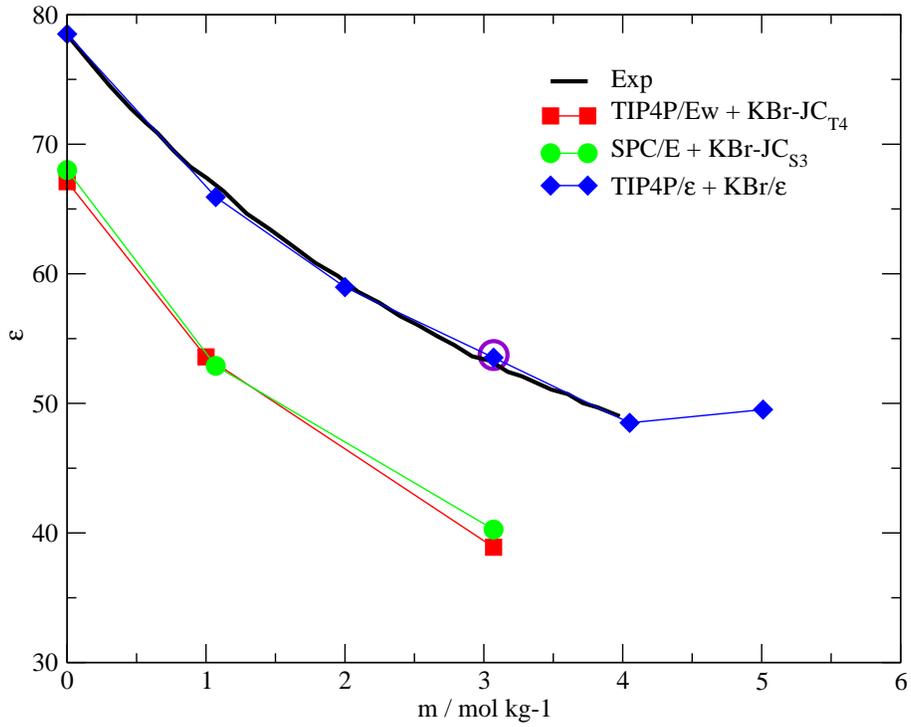}}
\caption{Dielectric constant versus molal
concentration of the KBr  salt at 1bar and 298 K. The black line is the experimental 
data~\cite{CRC}, the blue filled diamond is the results for the
KBr/$\varepsilon$ model , the green
spheres are the results for the JC$_{S3}$ model while the red 
squares are the results for the JC$_{T4}$ model.  The violet circle shows
 the concentration used for 
the parametrization of the model.}
\label{Fig4}
\end{figure}

The figure~\ref{Fig4} illustrates the 
the dielectric constant versus molal concentration
for the  KBr/$\varepsilon$ model with the TIP4P/$\varepsilon$ water, for the
experiments~\cite{CRC}, JC$_{S3}$~\cite{SD94} and JC$_{T4}$~\cite{JC08} models. The 
molal concentration imployed
to parametrize the KBr/$\varepsilon$ is shown with a purple circle. The result
shows that our model gives a good agreement with the experiments.

\begin{figure}
\centerline{\psfig{figure=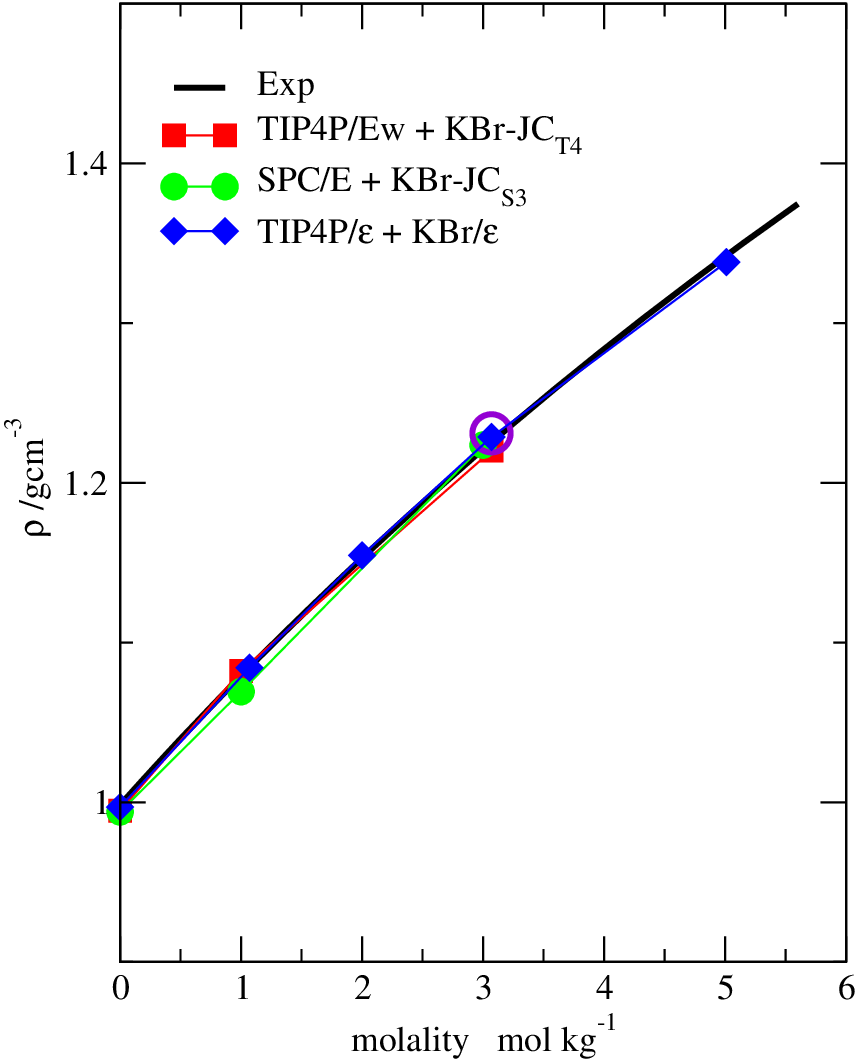,width=12.0cm,angle=0}}
\caption{ Density versus molal concentration of the 
salt  at 298 K
and 1bar.The black line is the experimental 
data~\cite{CRC}, the blue filled diamond is the results for the
KBr/$\varepsilon$ model , the green
spheres are the results for the JC$_{S3}$ model while the red 
squares are the results for the JC$_{T4}$ model. All 
data are  at 1bar and 298 K. The iolet circle shows the concentration used for 
the parametrization of the model.}
\label{Fig5}
\end{figure}

In the figure~\ref{Fig5} shows the density versus salt concentration
for the experiments~\cite{CRC}, for the 
KBr/$\varepsilon$, for the JC$_{S3}$~\cite{SD94} and for the JC$_{T4}$~\cite{JC08} models. 
The results are consistent with the fact that the models were parametrize to
give the correct density value.

\begin{figure}
\centerline{\psfig{figure=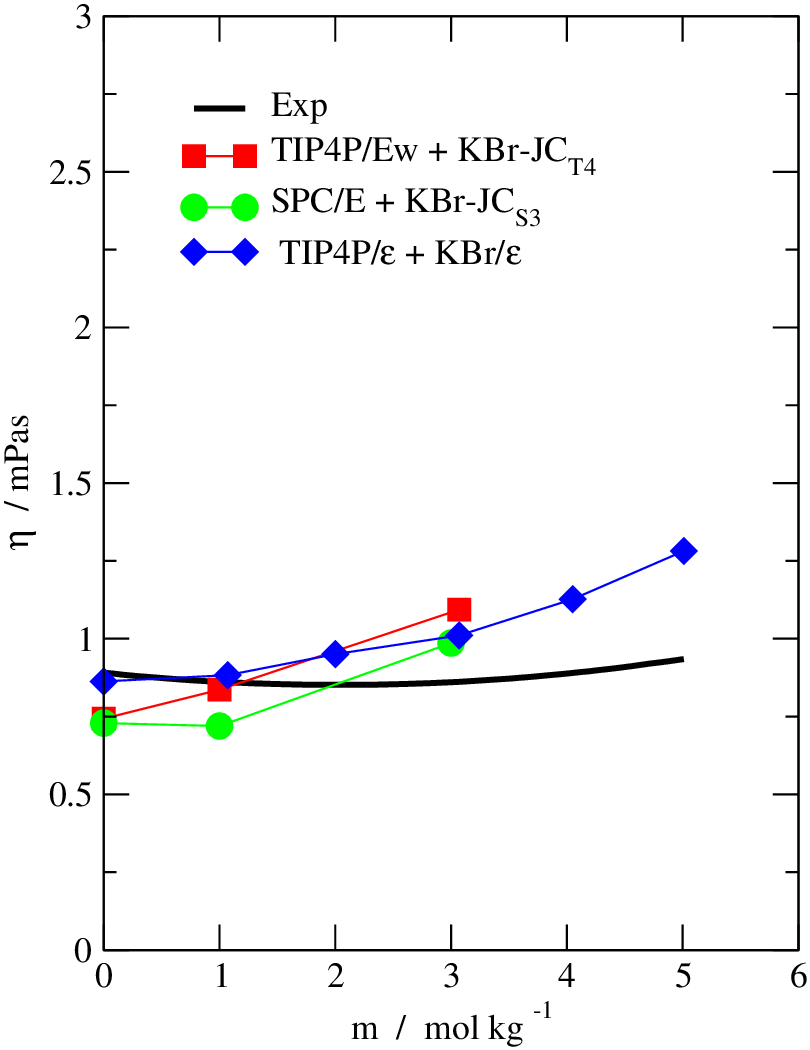,width=12.0cm,angle=0}}
\caption{Viscosity versus molal
concentration of the salt  at 298 K
and 1bar.The black
 line is the experimental data~\cite{CRC} 
and the blue filled diamond are the results for our model.}
\label{Fig6}
\end{figure}

In addition to thermodynamic properties, the 
transport was also evaluated. The
 shear viscosity $\eta$  at different 
molal concentrations at $1$  bar and $298\;K$ of pressure
 and temperature respectively. The figure~\ref{Fig6} illustrates
the  viscosity
versus the molal concentration of the salt showing 
an increase of $\eta$ as the salt concentration
increases what implies that the system becomes
more viscous. This result is consistent with
the experimental values~\cite{CRC} at diluted concentration 
also shown in the same figure. When the concentration 
is increased the
agreement with the experiments is lost.

\begin{table}[h]
\caption{Ion-Water Coordination Numbers 
obtained by our simulations  along 
 the r-range used in the integration.}
\label{ncoordKBr}
\begin{center}
\begin{tabular}{|c|cc|}
\hline
molal& MD & MD \\
 concentration	&	KO	&	BrO\\
\hline
\hline
3.07	&	5.38	&	5.13	\\
5	&	4.71	&	4.51 \\

\hline
\end{tabular}
\end{center}
\end{table}

\begin{figure}
\centerline{\psfig{figure=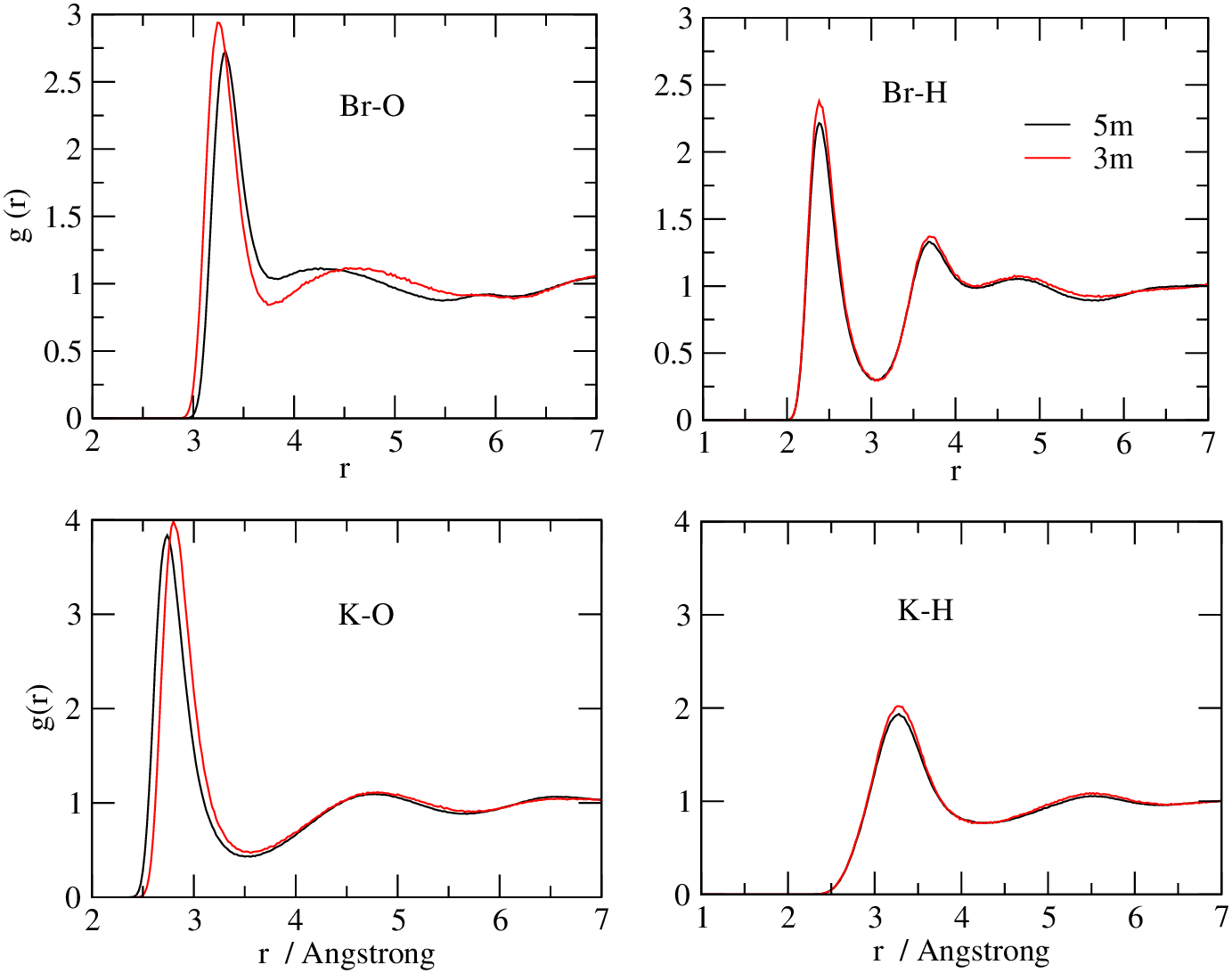,width=12.0cm,angle=0}}
\caption{Ion-water pair distribution functions using the rigid water model TIP4P/$\varepsilon$ and KBr/$\varepsilon$ force field at 298 K, 1 bar, and ionic concentrations of 5 (black line) and 3 (red line) molal. In all cases 864 molecules were used.}
\label{Fig-gr-KBr-h2o}
\end{figure}

The water coordination numbers around the K and Br
 ions,  \ref{ncoordKBr}, can be estimated by integrating the area under the first
 peak of the K-O and Br-O pair distribution functions up to
 the first minimum respectively, \ref{Fig-gr-KBr-h2o}.

\subsection{ The   KCl/$\varepsilon$ Model}

Since both NaCl/$\varepsilon$ and KBr/$\varepsilon$ models have been
already parametrized, we test the transferability of these two 
force fields as follows. Instead of fitting the parameters 
for  the KCl from experiments, the parameters
for the K are taken from the KBr/$\varepsilon$ given in the
Table~\ref{tab:KBrParam} while the parameters for the Cl
are taken from the NaCl/$\varepsilon$  model~\cite{NaCle}.
The parameters for the KCl are summarized 
in the  Table~\ref{tab:KClParam}. It is important to notice
that no additional parameterization was needed for 
obtaining the KCl/$\varepsilon$ model.

 \begin{table}[h]
\caption{Force field parameter of KCl/$\varepsilon$. }
\label{tab:KClParam}
\begin{center}
\begin{tabular}{|ccccc|}
\hline\hline
Model  & q/e&$\lambda_{C}$& 
$\sigma$/\AA & $(\epsilon/k_B)$/K\\
\hline
K&+1&0.885&2.86&115.83\\
Cl&-1&0.885&3.85  & 192.45 \\
\hline
\end{tabular}
\end{center}
\end{table}

\begin{figure}
\centerline{\psfig{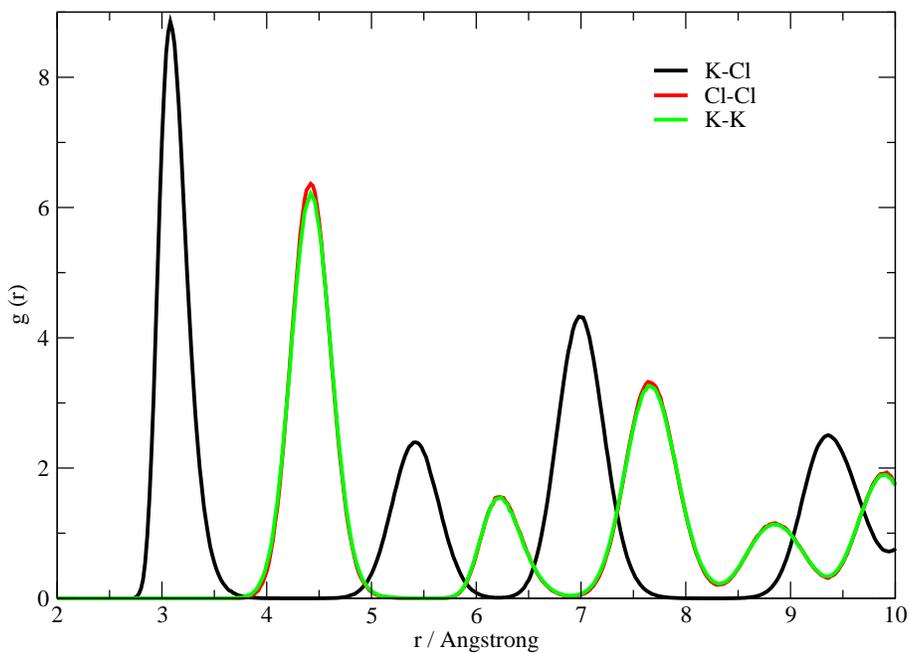}}
\caption{Radial distribution function g(r) versus the distance r at 1
bar and 298 K for: Cl-Cl (red line), K-K (green line), and K-Cl
(black line).
}
\label{Fig-g-r-KCl}
\end{figure}

First, we test the value of 
the density of the crystal of KCl/$\varepsilon$  
at the 298 K and 1 bar of pressure. Our
results give   $1.99 \;g \;cm^{-3}$, that is the same as the
experimental data~\cite{CRC}. The  radial distribution for
K-K, Cl-Cl and K-Cl is  illustrated
in the Figure~\ref{Fig-g-r-KCl} and it shows  a peak  at 
$3.08\AA$ in agreement with the experiments~\cite{CRC}. \\
\\
\\
\newpage
The Lattice
Constant (LC) at 1 bar of pressure and 298 K of temperature,  illustrated in 
the Table~\ref{tab:FFcompKCl}, is also in accordance with
the experimental values~\cite{CRC}.
The Lattice Energy (LE), however, it
is 16.5$\%$ far from the reproduction of the Lattice Energy.
This difference was present in the NaCl/$\varepsilon$ and KBr/$\varepsilon$ and
models.

\begin{table}[h]
\caption{Density of KCl at 1 bar of pressure and 298 K of temperature,
Lattice Energy, Lattice  Crystal 
of various force fields and for experiments~\cite{CRC}.}
\label{tab:FFcompKCl}
\begin{center}
\begin{tabular}{|cccc|}
\hline
 Model Ions& $\rho/(g/cm^3)$ & LC/\AA&LE/(kJ/mol) \\
\hline
JC$_{s3}$~\cite{SD94}& 1.86 &6.38&720.9\\
JC$_{T4}$~\cite{JC08} &1.90&6.34 &724.6\\

KBr/$\varepsilon$ &1.99 & 6.29&600.77\\
experimental~\cite{CRC}  &1.99&6.26&720.06\\
\hline
\end{tabular}
\end{center}
\end{table}

\begin{figure}
\centerline{\psfig{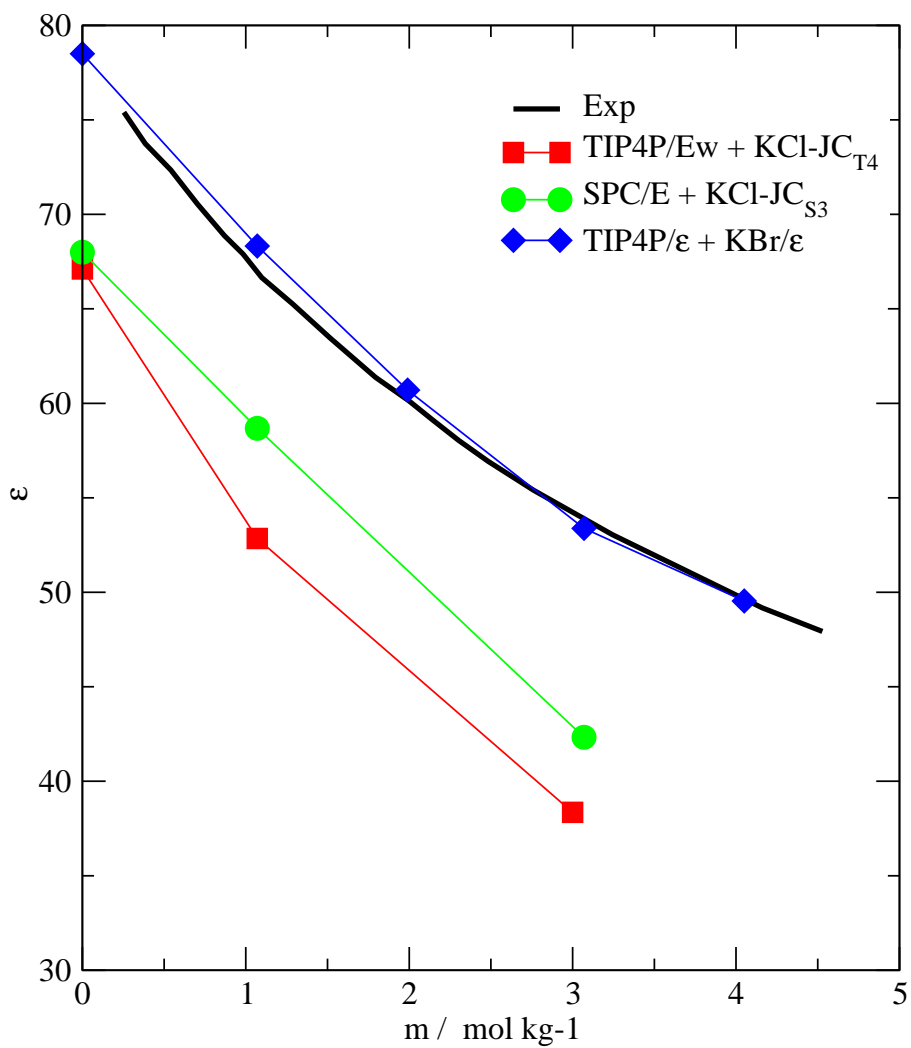}}
\caption{Dielectric constant versus molal
concentration of the salt at 298 K and 1bar . The black line 
is the experimental 
data~\cite{CRC} and the blue filled diamond 
is the results of our model. All 
data are  at room conditions. }
\label{Fig10}
\end{figure}

\begin{figure}
\centerline{\psfig{figure=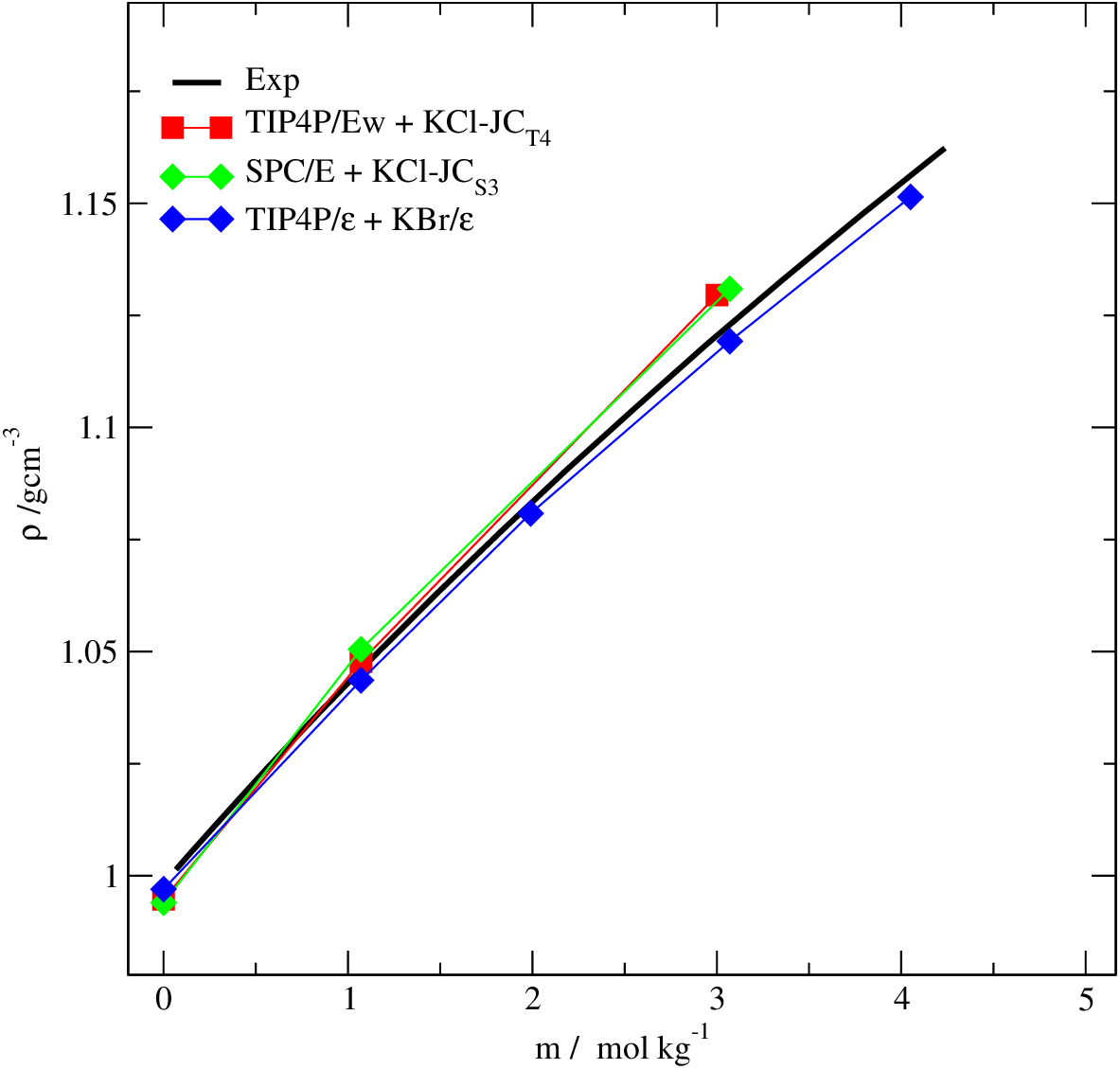,width=12.0cm,angle=0}}
\caption{ Density versus molal concentration of the 
salt  at 298 K and 1bar. The black line is the experimental data~\cite{CRC} 
and the blue filled diamond are the results of 
our model. }
\label{Fig11}
\end{figure}

Next, the  KCl/$\varepsilon$ in the water TIP4P/$\varepsilon$ performance is
tested  for a number 
of properties. The dielectric constant of the solution 
as a function of the  molal concentrations is shown 
in the figure~\ref{Fig10} giving a good agreement with
the experiments~\cite{CRC}. Similarly  the figure~\ref{Fig11}
gives the density at a function
of the molal concentration of KCl. 
 
\begin{figure}
\centerline{\psfig{figure=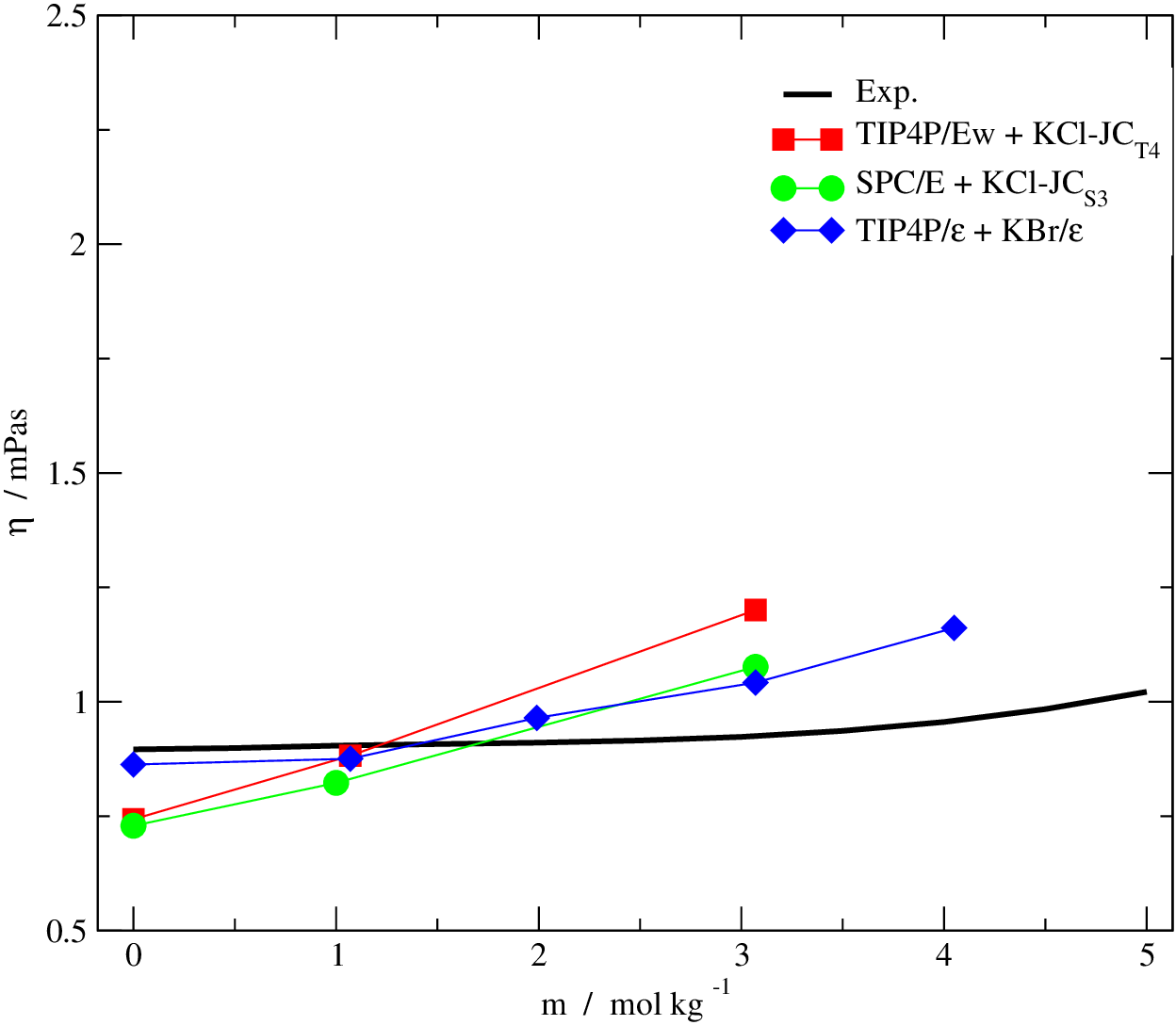,width=12.0cm,angle=0}}
\caption{Viscosity versus molal
concentration of the salt
at temperature and pressure at room conditions.The black
 line is the experimental data~\cite{CRC} 
and the blue filled diamond are the results for our model.}
\label{Fig12}
\end{figure}

The  shear viscosity $\eta$  at different 
molal concentrations at 1 bar and 298K  is
shown in the figure~\ref{Fig12} showing a 
good agreement with the experiments.

\begin{figure}
\centerline{\psfig{figure=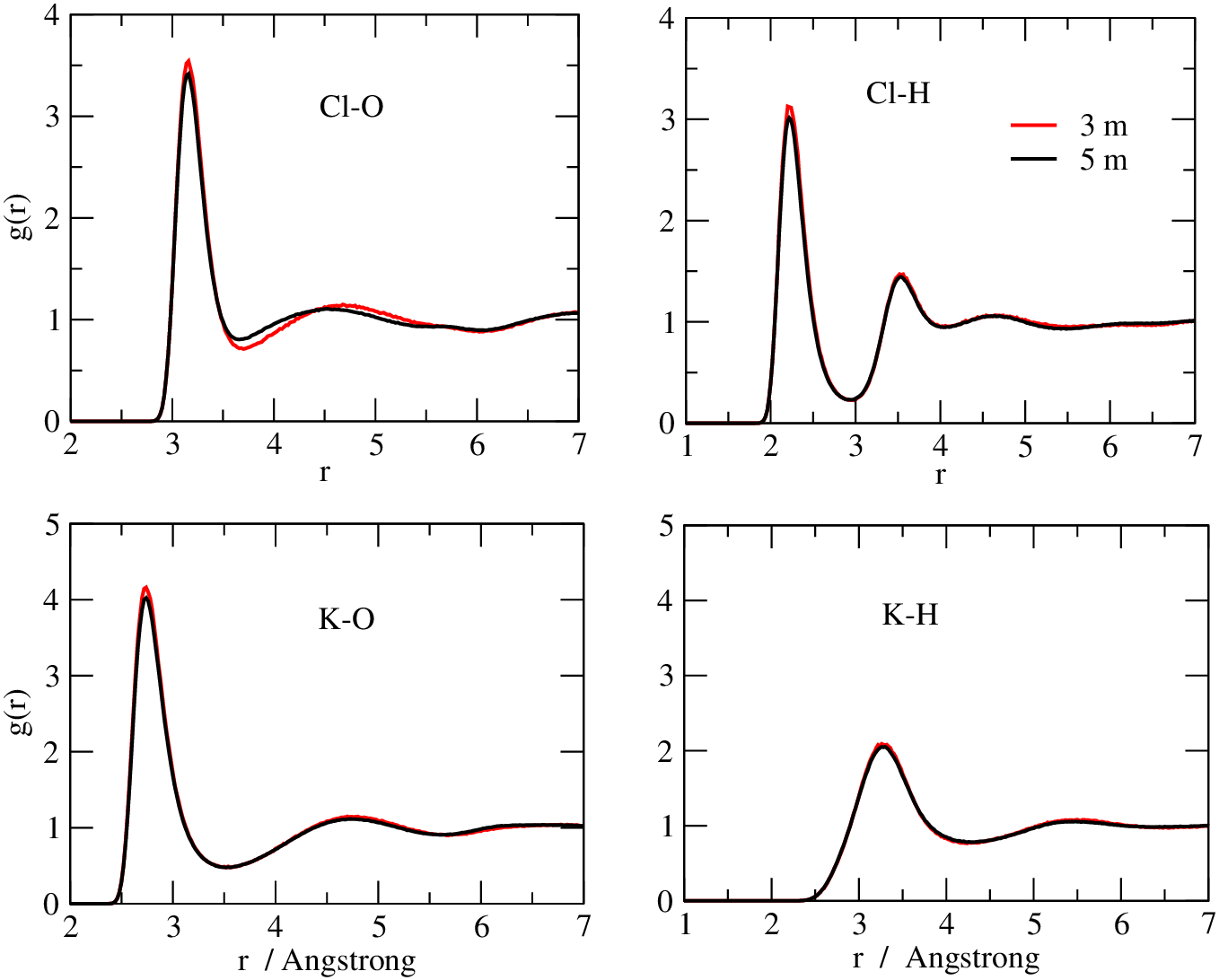,clip,width=12.0cm,angle=0}}
\caption{Ion-water pair distribution functions using the rigid water model TIP4P/$\varepsilon$ and KCl/$\varepsilon$ force field at 298 K, 1 bar, and ionic concentrations of 5 (black line) and 3 (red line) molal. In all cases 864 molecules were used.}
\label{Fig-gr-KCl-h2o}
\end{figure}
Using the radial distribution functions respect to oxygen of the cation and anion at different concentrations. We calculated the number of coordination around the K and Cl, we do this through the integration the area under the first
 peak of the K-O and Cl-O pair distribution functions up to
 the first minimum respectively, \ref{Fig-gr-KCl-h2o}. These coordination numbers 
are shown in the table~\ref{ncoordKCl} and give a good
agreement with the experiments in the case of KO.

\begin{table}[h]
\caption{Ion-Water Coordination Numbers 
obtained by our simulations and experiments.  The uncertainties of 
experimental data~\cite{Mancinelli} are reported within parenthesis, along 
with the r-range used in the integration.}
\label{ncoordKCl}
\begin{center}
\begin{tabular}{|c|cc|c|}
\hline
molal& MD & MD &Exp~\cite{Yizhak}  \\
 concentration	&	KO	&	ClO&KO	\\
\hline
\hline
3.07	&	5.61	&	5.46	& 5.7 \\
5	&	5.18	&	5.10 & 5.1\\

\hline
\end{tabular}
\end{center}
\end{table}



\subsection*{The NaBr/$\varepsilon$ Model}

\begin{figure}
\centerline{\psfig{figure=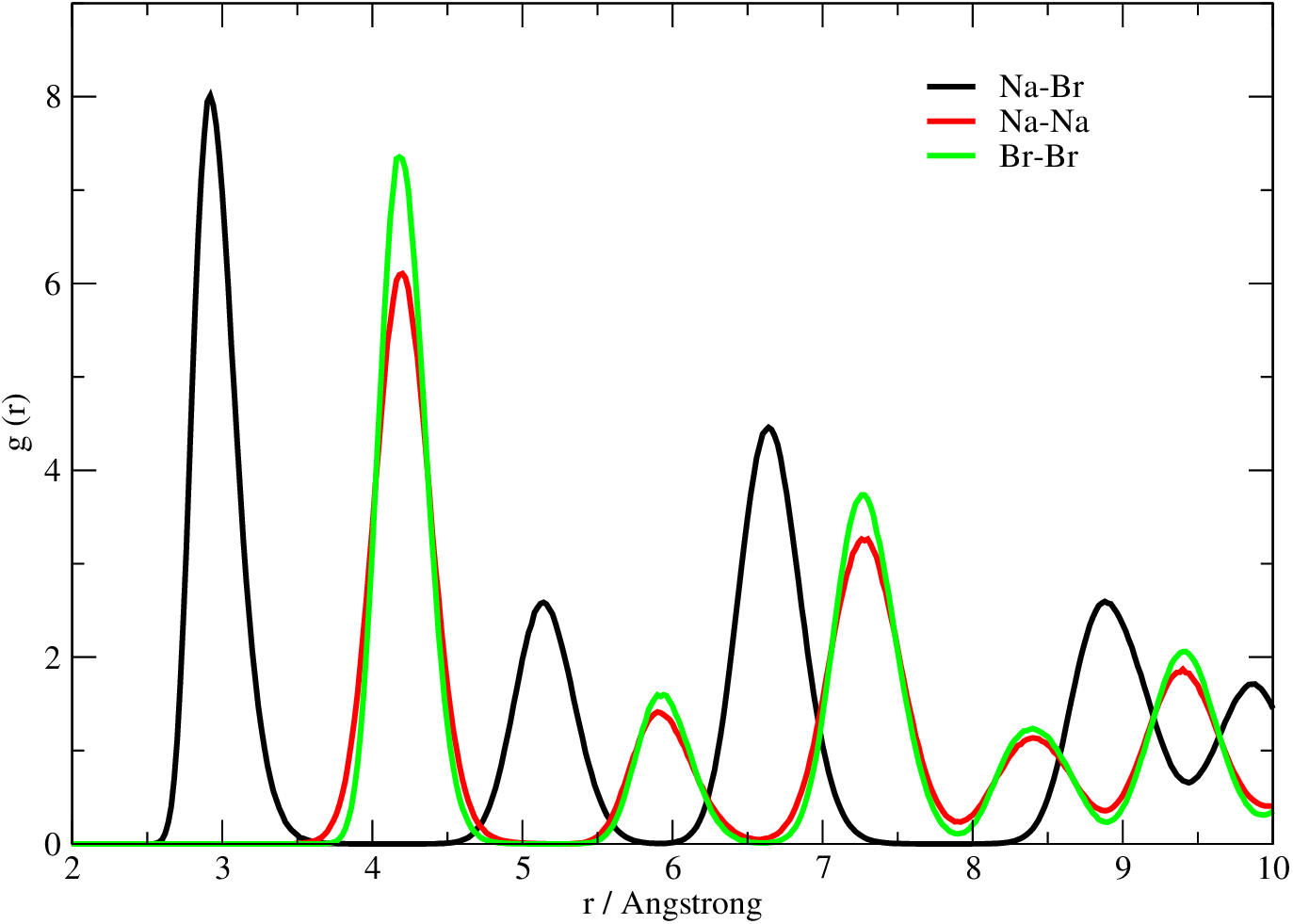,width=12.0cm,angle=0}}
\caption{Radial distribution function g(r) versus the distance r at 1
bar and 298 K for: Na-Na (red line), Br-Br (green line), and Na-Br
(black line).
}
\label{Fig-g-r-NaBr}
\end{figure}

The consistency of the
new  force fields is
now checked by creating the   NaBr/$\varepsilon$ model
employing the parameters for Br and Na, shown
in the table~\ref{tab:NaBrParam}, extracted from the force fields for
the
 KBr/$\varepsilon$ and  NaCl/$\varepsilon$ models respectively. 
 The  radial distribution for
Na-Na, Br-Br and Na-Br are  illustrated
in the figure~\ref{Fig-g-r-NaBr}. The density of
the crystal, the lattice  energy and the lattice
constant are shown in the table~{tab:FFcompNaBr}
showing a good agreement with the experimental results.

 \begin{table}[h]
\caption{Force field parameter of NaBr/$\varepsilon$. }
\label{tab:NaBrParam}
\begin{center}
\begin{tabular}{|ccccc|}
\hline\hline
Model  & q/e&$\lambda_{C}$& 
$\sigma$/\AA & $(\epsilon/k_B)$/K\\
\hline
Na&+1&0.885&2.52 &17.44 \\
Br&-1&0.885&4.057&287.47\\
\hline
\end{tabular}
\end{center}
\end{table}
\newpage

\begin{table}[h]
\caption{Density of NaBr at 1 bar of pressure and 298 K of temperature,
Lattice Energy, Lattice  Crystal 
of various force fields and for experiments~\cite{CRC}.}
\label{tab:FFcompNaBr}
\begin{center}
\begin{tabular}{|cccc|}
\hline
 Model Ions& $\rho/(g/cm^3)$ & LC/\AA&LE/(kJ/mol) \\
\hline
JC$_{s3}$~\cite{SD94}& 3.00 &6.06&761.48\\
JC$_{T4}$~\cite{JC08} &3.09&6.00 &766.50\\

NaBr/$\varepsilon$ &3.2 & 5.90&600.77\\
experimental~\cite{CRC}  &3.2&5.97&753.95\\
\hline
\end{tabular}
\end{center}
\end{table}

\begin{figure}
\centerline{\psfig{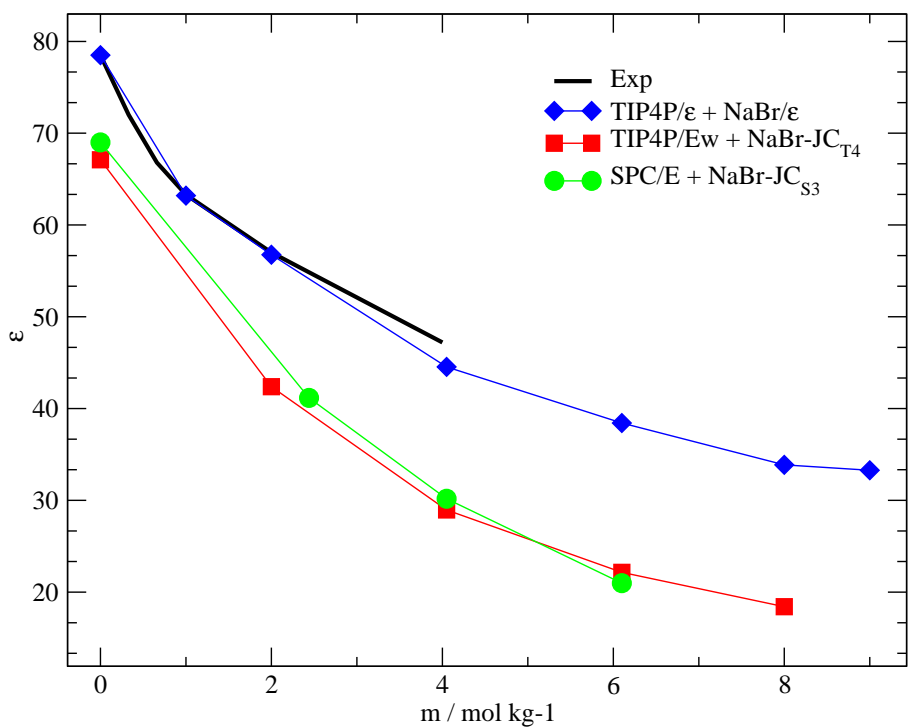}}
\caption{Dielectric constant versus molal
concentration of the salt. The black line is the experimental 
data~\cite{CRC} and the blue filled diamond 
is the results of our model. All 
data are  at room conditions. Violet circle is the diluted concentration where was made the parameterization.}
\label{Fig16}
\end{figure}

The  figure~\ref{Fig16} illustrates
the dielectric constant versus molal concentration
of the  NaBr/$\varepsilon$ salt model in solution
with the  TIP4P/$\varepsilon$ water. The graphs shows
that the new model gives a better agreement
with the experiments than the other atomistic parameterizations. 

\begin{figure}
\centerline{\psfig{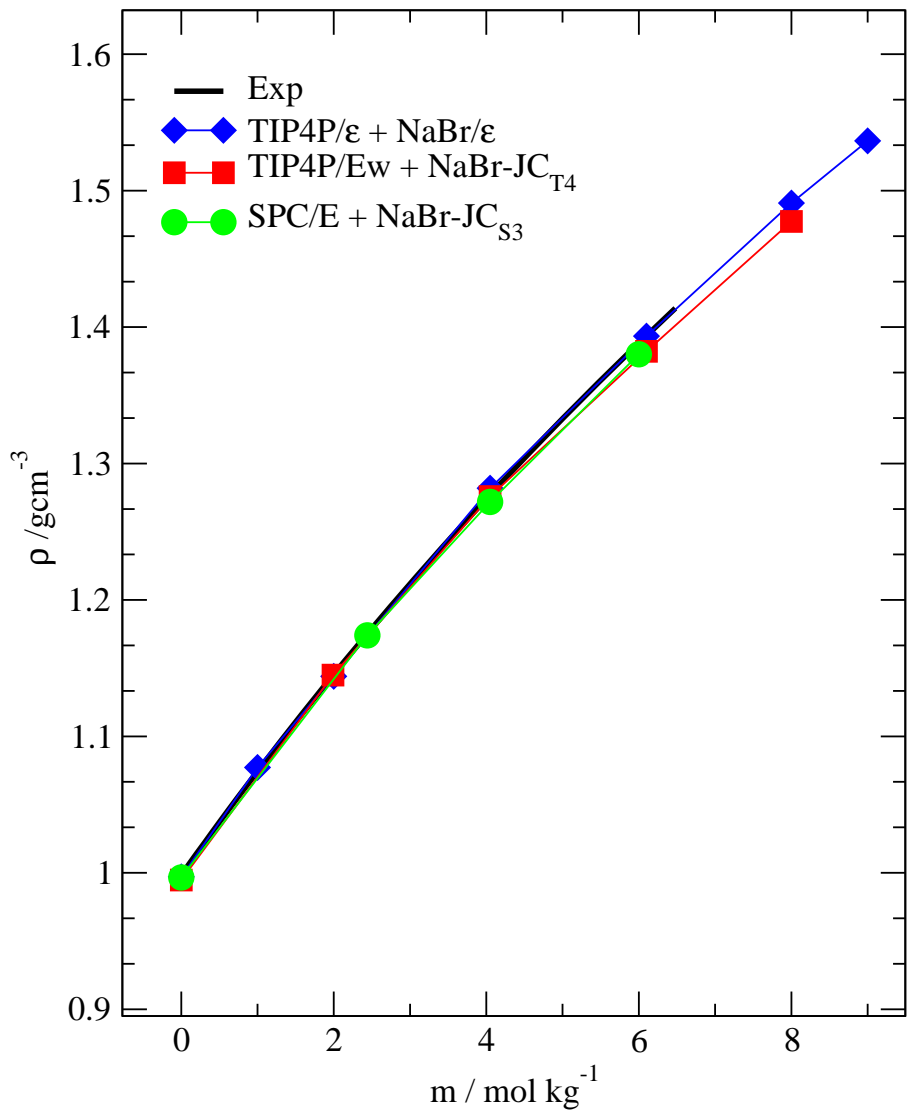}}
\caption{ Density versus molal concentration of the 
salt  at 298 K
and pressure.The black line is the experimental data~\cite{CRC} 
and the blue filled diamond are the results of our model. Violet 
circle is the diluted concentration where was made the parameterization.}
\label{Fig17}
\end{figure}

The density of the  NaBr/$\varepsilon$  with the TIP4P/$\varepsilon$ water
 at various concentrations is show in the figure~\ref{Fig17}. These force 
fields reproduce the density of the diluted solution very well and 
when the solution increases in concentration we see that the value 
the same with respect to experimental, so in a 6m solution we
 find that the value of the JC$_{S3}$ is 1.5 percent and the 
JC$_{T4}$ is 1.6 percent diferent with respect to the experimental.

\begin{figure}
\centerline{\psfig{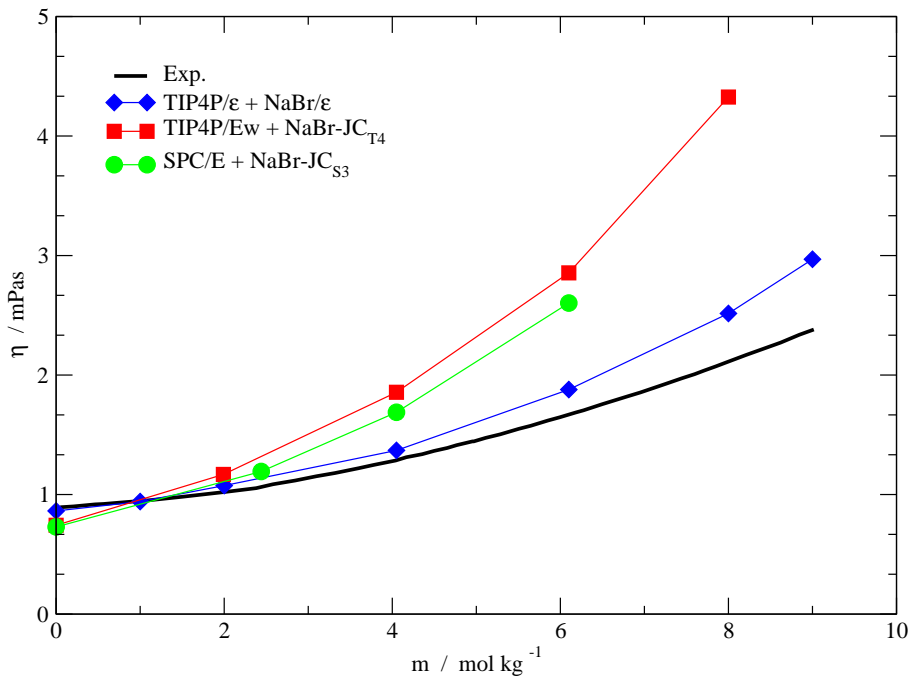}}
\caption{Viscosity versus molal
concentration of the salt
at temperature and pressure at room conditions.The black
 line is the experimental data~\cite{CRC} 
and the blue filled diamond are the results for our model.}
\label{Fig18}
\end{figure}

\newpage
The calculus of the  shear viscosity $\eta$  at different 
molal concentrations at 1 bar and 298K of pressure and 
temperature respectively, \ref{Fig18} show the shear viscosity
versus molal concentration of the salt showing 
an increase of $\eta$ as the salt concentration
increases what implies that the system becomes
more viscous. This result is consistent with
the experimental values~\cite{CRC} at diluted concentration also shown 
in the figure. When the concentration is increase there is a 
difference between the value that reproduces the NaBr/$\varepsilon$ 
with TIP4P/$\varepsilon$ of 13.5$\%$ at 6.1m, compared to experimental
 value and the JC$_{S3}$ with respect to the experimental value 
is 55.3 percent and the JC$_{T4}$ is 71.4 percent at the 
same experimental concentration of 6.1m. It is important to
 note that the result with the new force field NaBr/$\varepsilon$ 
describes the curvature of the shear viscosity.\\
\\

\begin{figure}
\centerline{\psfig{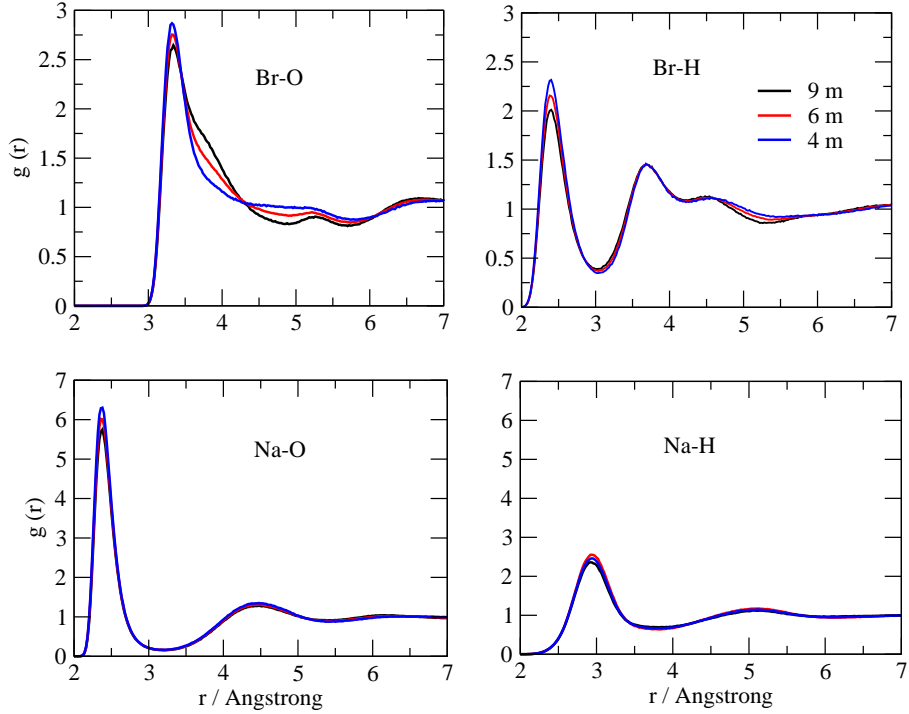}}
\caption{The radial distribution function for (a) Br-O, (b) Br-H, 
(c) Na-O and (d) Na-H in 
TIP4P/$\varepsilon$ water with  NaBr/$\varepsilon$
at salt concentrations of  9 (black line), 6 (red line) and
4 (blue line)
molal. In all cases 864 molecules were employed.}
\label{fig:gr-NaBr-h2o}
\end{figure}
\newpage
The ion-water radial distribution functions
are illustrated in the figure~\ref{fig:gr-NaBr-h2o}. Employing
this functions, the coordination around the NA and Br were
computed and shown in the table~\ref{ncoordNaBr}.

\begin{table}[h]
\caption{Ion-Water Coordination Numbers 
obtained by our simulations.}
\label{ncoordNaBr}
\begin{center}
\begin{tabular}{|c|cc|}
\hline
molal& MD & MD \\
 concentration	&	NaO	&	BrO\\
\hline
\hline
4	&	4.62	&	13.8	 \\
6	&	4.14	&	14.11 \\
9	&	3.65	&	12.86 \\
\hline
\end{tabular}
\end{center}
\end{table}

\begin{figure}
\centerline{\psfig{figure=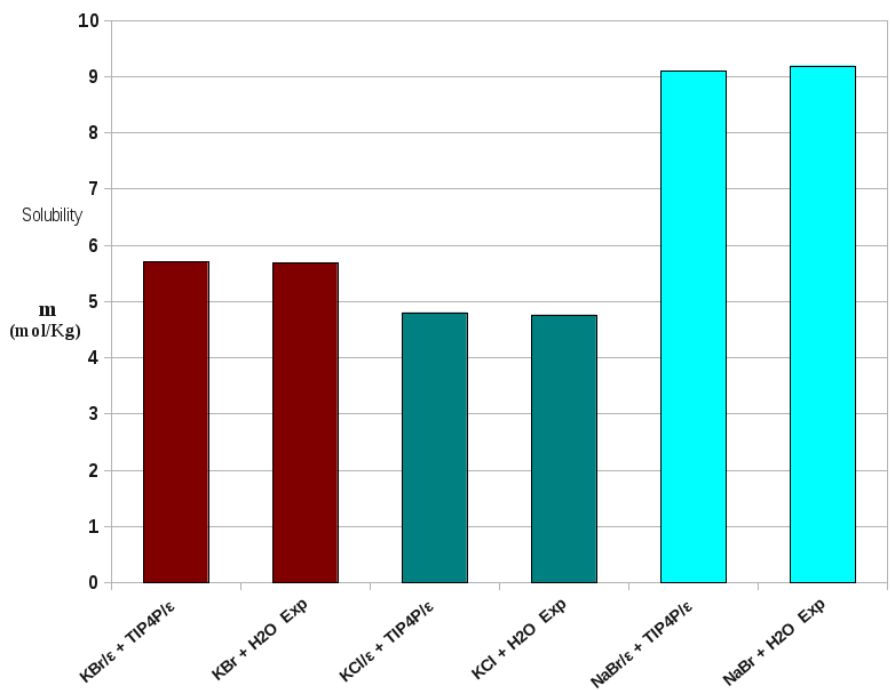,width=12.0cm,angle=0}}
\caption{Solubility of the atomistic models }
\label{solub}
\end{figure}

In the figure~\ref{solub} the solubilities of the new force fields 
 KBr/$\varepsilon$, KCl/$\varepsilon$ and 
NaBr/$\varepsilon$ are shown 
indicating a good agreement with
the experiments.The solubility was computed employing method number
four from the reference by Manzanilla-Granados et al.~\cite{Hector}

\section{Conclusions}

In this paper the force field,
the  KBr/$\varepsilon$, was introduced.
The model reproduces the density
of the crystal and structure, as well as the thermodynamic
and dynamic properties of the
solution with the   TIP4P/$\varepsilon$ model for water at different molal concentration.

The model, particularly reproduces the dielectric constant
of the solution what is a property not well represented
in other atomistic models.
In order to test the KBr/$\varepsilon$ and the NaCl/$\varepsilon$
models, the same parameters for the isolated ions were employed
in the construction of the KCl/$\varepsilon$ and of 
the NaBr/$\varepsilon$ models. In the construction
of these two new models not there was an additional parameterization.  The results of the  density, the  dielectric
constant, the viscosity and the solubility of the  KCl/$\varepsilon$ and of 
the NaBr/$\varepsilon$ models reproduces well the experiments. Since
 all the models give a very robust result for the dielectric constant, 
we believe that they are suitable for studying confined systems and can be used to study ionic channels.

\section {Acknowledgements}
 
 We thank the Brazilian agencies CNPq, INCT-FCx, and Capes for the
financial support. We also thank the SECITI of Mexico city for financial support.  

R.F.A. thanks Jos\'e Alejandre Ramirez chief of the Chemistry Department at UAM-Iztapalapa in Mexico city for all the advice and helpful discussions.

\end{document}